# Improving spatial resolution in X-ray microscopy by using tilted angle detector: A simulation study

*Polad Shikhaliev and Nicola Tartoni*

*Detector Group, Diamond Light Source,*
*Rutherford Appleton Laboratory, Harwell Science & Innovation Campus*
*Didcot, Oxfordshire, UK*

## ABSTRACT

Conventional x-ray imaging detectors suffer from parallax error when the radiation beam arrives at the detector surface at tilted angle. The image blurring occurs as the radiation penetrates detector material in lateral direction at tilted angle. However, at the photon energies used in x-ray microscopy the attenuation length of the x-rays in the detector material is shorter and consequently the parallax error due to the tilted angle irradiation can be smaller. In these cases, the tilted angle irradiation can be beneficial because it will expand the beam projection ("footprint") at the detector surface, improve spatial resolution and help to overcome inherent resolution limit of the detector. We have performed a simulation study to investigate the above tradeoff between the parallax error and improved spatial resolution when tilted angle irradiation is used. Our simulation study showed that tilted angle irradiation at 2-5keV x-ray energies can provide substantial improvements of spatial resolution. We used in our studies the indirect-conversion x-ray microscopy detectors based on scintillators optically coupled to CCD and CMOS cameras, as well as direct-conversion detectors based on back illuminated (back thinned) CCD and CMOS cameras. We have demonstrated also that the tilted angle irradiation can prevent radiation damage of the direct conversion back illuminated CCD and CMOS cameras due to the shallow x-ray attenuation depths with tilted angle geometry.

Corresponding author:

Polad Shikhaliev, PhD
Detector Group
Diamond Light Source
Rutherford Appleton Laboratory
Harwell Science & Innovation Campus
Didcot, Oxfordshire, UK
polad.shikhaliev@diamond.ac.uk

## 1. Introduction

X-ray imaging detectors exhibit best performance when they are irradiated perpendicular to the surface. Any deviation of the incidence angle from perpendicular (normal) may result in image blurring due to the parallax effect. When the x-ray beam arrives at tilted angle it may penetrate the detector in lateral direction and smear the true position of the detected photons. The negative effect of the tilted angle irradiation is well known in many areas. In medical radiography, for example, this effect decreases the spatial resolution over the peripheries of the detector where the beam arrives at less than normal angle (Webb 1988; Hasegawa 1991; Que and Rowlands 1995; Beutel et al. 2000; Badano et al. 2011). Similarly, in positron emission tomography (PET), the tilted angle incidence of the annihilation photons results in image blurring and quantification errors due to the parallax effect (Webb 1988; Cherry et al. 2003).

However, when the x-ray energy is decreased, the magnitude of the parallax error due to tilted angle irradiation is also decreased. At lower photon energies used in x-ray microscopy the attenuation length of the x-rays in the detector material can be shorter and parallax error can be smaller. On the other hand, tilted angle irradiation expands the beam projection ("footprint") at the detector surface, which allows for increasing spatial resolution and overcoming intrinsic resolution limits of the detector. Finding balance between these two conflicting effects, parallax error and improved spatial resolution with a tilted angle detector is the subject of this paper.

The results of our simulation study demonstrated the possibility of substantial improvements of spatial resolution when the low energy x-rays are used in tilted angle irradiation geometry. For example, the tilted angle irradiation increased spatial resolution by a factor of 5 for 2-5keV x-rays as compared to normal irradiation. The tilted angle irradiation can be applied for different types of detectors. In the current work we considered two detector systems that are widely used for x-ray microscopy (Withers 2007; Sakdinawat and Attwood 2010; Kaulich et al. 2011). The first is an indirect-conversion detector based on scintillator optically coupled to an imaging camera such as CCD or CMOS. The second detector is a direct-conversion back-illuminated (back-thinned) CCD or CMOS detector that includes a depleted thin Si layer serving as x-ray converter. The tilted angle irradiation provides also several other advantages, such as prevention of the back-illuminated (back-thinned) CCD and CMOS detectors from radiation damage due to shallow x-ray absorption depths. These and other aspects of the method are described in details along with possibilities of its practical realization.

## 2. Methods and materials

### 2.1. Concept of tilted angle irradiation

The tilted angle irradiation of the detectors was previously investigated for various purposes. For example, Si strip detectors used for medical x-ray imaging were irradiated at small tilt angle (i.e., the angle between the detector surface and x-ray beam) to increase absorption efficiency of 15-45keV x-rays (Lundqvist et al. 2001). In this case the parallax effect did not create a problem because Si strip detector was a 1D detector and its signal electrodes were along the x-ray beam direction. The tilted angle geometry was studied for linear array CdZnTe detectors for computed tomography (CT) applications to increase detector count rate (Shikhaliev 2006). The tilted angle irradiation of CdZnTe detector was also used for x-ray spectroscopy where it provided substantial improvement on low energy tailing of the energy spectrum due to hole trapping (Fritz and Shikhaliev 2009; Fritz et al. 2011).

It should be noticed that the tilted angle concept was previously used also for improving spatial resolution of synchrotron beam-based x-ray microscopy (Sakamoto et al. 1988; Silver 1995; Schafer and Kohler 2003; Honnicke and Cusatis 2007). However, in these works the x-ray beam was expanded by a tilted angle x-ray mirrors not by the detector. In these studies the beam expanding mirror was installed between the imaged object and detector.

In the current work we investigated possibilities of improving spatial resolution by tilted angle irradiation of the detector. In **figure 1** two pencil beams of x-rays with small gap in between arrive at the detector surface at normal (90º) angle and tilted angles, interact with the detector, and two signals are generated (the angles here and further are measured between the detector surface and x-ray beam).

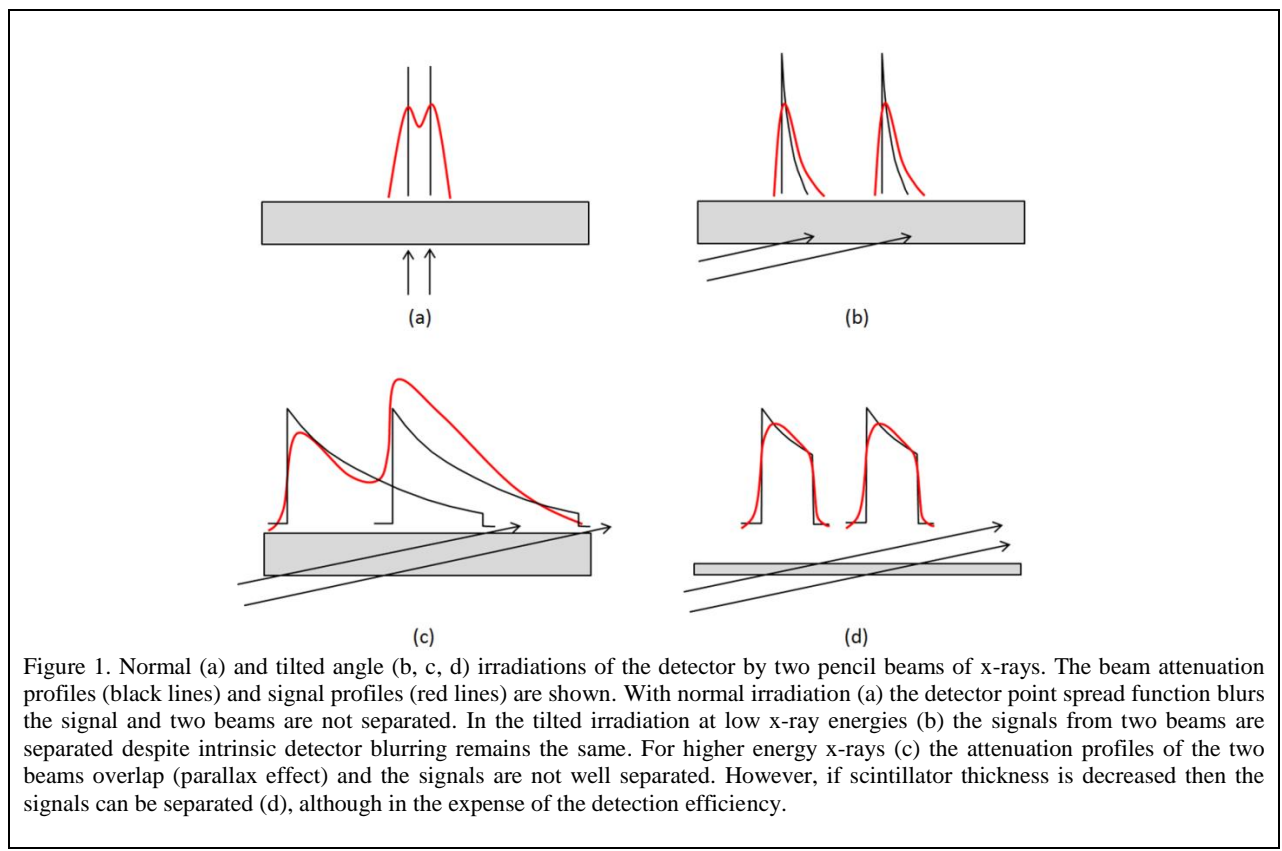

Figure 1. Normal (a) and tilted angle (b, c, d) irradiations of the detector by two pencil beams of x-rays. The beam attenuation profiles (black lines) and signal profiles (red lines) are shown. With normal irradiation (a) the detector point spread function blurs the signal and two beams are not separated. In the tilted irradiation at low x-ray energies (b) the signals from two beams are separated despite intrinsic detector blurring remains the same. For higher energy x-rays (c) the attenuation profiles of the two beams overlap (parallax effect) and the signals are not well separated. However, if scintillator thickness is decreased then the signals can be separated (d), although in the expense of the detection efficiency.

The shape and width of the signals are determined by the detector point spread function (PSF) and attenuation profiles of the x-ray beams. In the case of normal irradiation (**figure 1a**) two signals are not separated because the distance between the beams is comparable with FWHM (full width at half maximum) of the detector PSF. In the second case (**figure 1b**) the beams arrive at small tilt angle, the x-ray energies are low and x-ray absorption length is short. The attenuation profiles and signals of the two beams are physically separated. When the x-ray energy is increased (**figure 1c**), the absorption length of the photons in the detector material is also increased, and the photon attenuation profiles of the two beams overlap creating parallax effect. In this case the signals from two beams are not fully separated. This problem can be addressed by decreasing scintillator thickness (**figure 1d**), although the detection efficiency would also decrease.

### 2.2. Study description

#### *2.2.1. Tilted angle scintillation detector*

Schematic of the indirect-conversion, scintillation detector system used in x-ray microscopy is shown in **figure 2**. It includes a thin layer of scintillator built on a transparent substrate, optically coupled to an imaging camera via an optical lens (Koch et al. 1998; Martin et al. 2009). F**igure 2a** shows the normal irradiation and the **figure 2b, c** show tilted angle irradiation geometries. The tilted angle irradiation can be performed in two configurations. In one configuration (**figure 2b**) the detector that is used in normal irradiation is simply turned by 90º minus tilt angle. In the second configuration (**figure 2c**) the scintillator with substrate is flipped so that the light is collected from scintillator directly without passing through substrate which allows for minimizing spherical aberration.

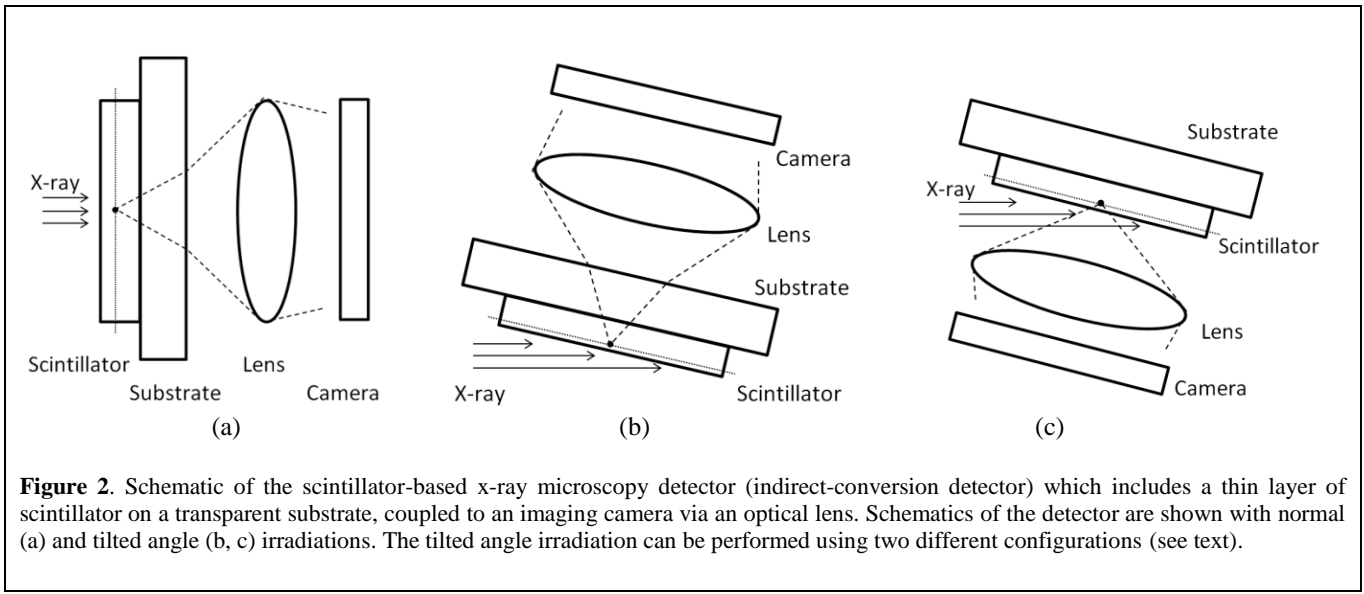

**Figure 2**. Schematic of the scintillator-based x-ray microscopy detector (indirect-conversion detector) which includes a thin layer of scintillator on a transparent substrate, coupled to an imaging camera via an optical lens. Schematics of the detector are shown with normal (a) and tilted angle (b, c) irradiations. The tilted angle irradiation can be performed using two different configurations (see text).

The x-ray beam passed through the imaged object irradiates the small area of the scintillator called field of view (FOV) which has a size in the order of 1x1mm$^2$. The scintillator typically has 1-20 micron thickness, and the thickness of the substrate can be 100-500 micron. The x-rays arriving at the scintillator are converted to light, and the light distribution (i.e., image) created within the FOV is magnified by 10-40 times by the optical lens. This magnified image is then projected to the surface of the optical image camera based on CCD or CMOS with typically 2048x2048 array of pixels with 5-15µm pixel size.

*a. Spatial resolution of the system*

Ideally, the scintillator layer where the light is generated should be infinitely thin and the optical lens should be focused to this thin layer. However, x-rays are attenuated at different depths and some part of the light is generated outside of the focus of the lens. This effect is called defect of focus and it results in blurring of the images (Koch et al. 1998). The magnitude of the blurring due to the defect of focus depends on numerical aperture ($NA$) of the lens and thickness $\Delta d$ of the scintillator layer where light is generated as $R_1 \sim \Delta d NA$. Another factor limiting spatial resolution is the diffraction of the light which limits the spatial resolution at the level comparable to the wavelength of the light. The blurring associated with this factor is determined as $R_2 \sim \lambda/NA$ where $\lambda$ is the wavelength of the scintillation light. The third factor contributing to the resolution blurring is spherical aberration of the lens which depends on substrate thickness and numerical aperture as $R_3 \sim t(NA)^3$, assuming that thickness of the scintillation layer is much smaller than the substrate thickness. Finally, the size of the pixel array of CCD/CMOS camera also affects the resolution. The lens magnification and geometry are selected such that the FOV at the scintillator is projected to whole surface of the CCD/CMOS camera. Therefore, the 2048x2048 pixel array of the camera samples the 1x1mm² FOV of scintillator providing 0.5µm effective pixel size at the scintillator. All four components of the resolution blurring are combined at the scintillator surface resulting in the total image blurring.

The work (Koch et al. 1998) performed comprehensive simulations and experimental studies of the above effects and derived following expression for the spatial resolution:

$$R_{50\%} = \sqrt{\left(\frac{0.18}{NA}\right)^2 + (0.075 d NA)^2} \qquad (1)$$

where $R_{50\%}$ is the FWHM broadening of the spatial resolution due to the defect of focus and diffraction limit, and $d$ is the thickness of the scintillator. Both the $R_{50\%}$ and $d$ are expressed in $\mu m$ and $NA$ is unitless. It was assumed that the light is created uniformly in the scintillator so that $\Delta d \equiv d$. The spherical aberration was not included in the expression (1) because it generally can be corrected by adding additional lens that corrects aberrations. It is noticed also that for the lenses with large $NA$ typically used in x-ray microscopy satisfactory correction for spherical aberration can be difficult to achieve (Koch et al. 1998).

There is another component of the blurring of spatial resolution associated with absorption lengths of the x-ray interaction products such as photo- and Auger-electrons and fluorescent photons. However, the Monte Carlo simulations and estimations using range data (NIST) showed that the magnitude of this blurring is below 0.1µm for the photon energies used in x-ray microscopy (Koch et al. 1998).

When tilted angle irradiation is used the spatial resolution is increased in 1D due to the geometric expansion of the beam cross-section. While the image of the object is also expanded in 1D along with the beam cross section, the detector pixel size and FWHM blurring remains the same (**figure 3**). The expanded image is then scaled back (demagnification) to its original size. In this demagnification the effective detector pixel size and FWHM detector blurring are decreased respectively improving spatial resolution. For 6º tilt angle used in the current work the beam is expanded by 10 times, and effective pixel size and detector blurring is decreased by 10 times, respectively.

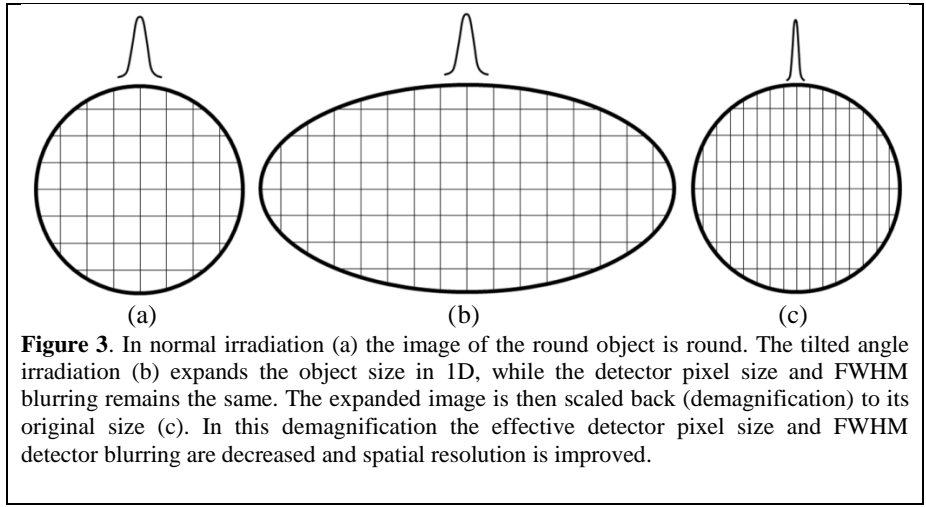

**Figure 3**. In normal irradiation (a) the image of the round object is round. The tilted angle irradiation (b) expands the object size in 1D, while the detector pixel size and FWHM blurring remains the same. The expanded image is then scaled back (demagnification) to its original size (c). In this demagnification the effective detector pixel size and FWHM detector blurring are decreased and spatial resolution is improved.

*b. Simulation parameters*

Our simulation study used $Lu_3Al_5O_{12}$ scintillator (called also LAG:Ce or LuAG:Ce) with 5µm thickness. This scintillator has high density of 6.73g/cm$^3$ and high effective atomic number of 63, and it is widely used in x-ray microscopy (Martin and Koch 2006; Martin et al. 2009). The attenuation length of LuAG:Ce scintillator is shown in **figure 4**.

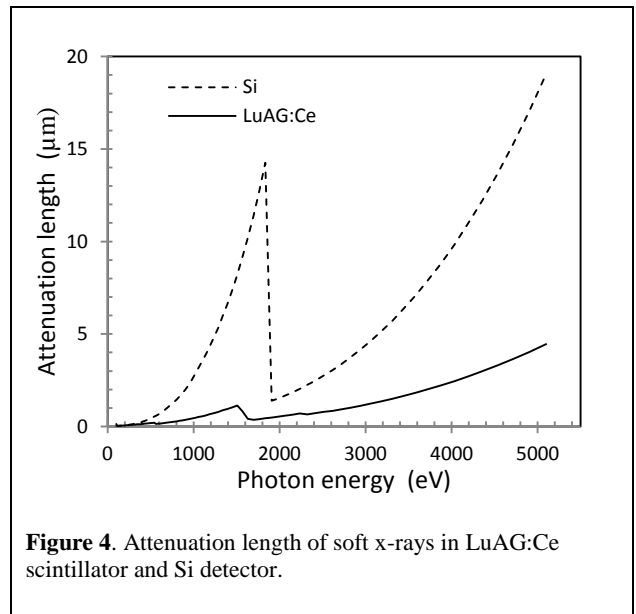

**Figure 4**. Attenuation length of soft x-rays in LuAG:Ce scintillator and Si detector.

The numerical aperture of the lens was assumed to be $NA = 0.75$ which is used in high resolution x-ray microscopy systems (Martin and Koch 2006). The FOV measured at the scintillator surface was 0.4x0.4mm$^2$ and the pixel array of the camera was 2048x2048, which corresponds to 0.2 µm effective pixel size in the FOV. With the above parameters the theoretically predicted broadening of the FWHM spatial resolution was 0.42 µm which includes broadenings due to the diffraction, defect of focus, and detector pixel size. However, the comparisons between the theoretical and measured broadenings show that the measured broadening is by a factor of 1.5- 2 larger than the calculated one (Koch et al. 1998). Therefore, we used in our simulations a realistic FWHM spatial broadening of 0.7 µm experimentally determined in (Koch et al. 1998) instead of theoretically predicted 0.42 µm.

The parameters used in simulations of the tilted angle illumination are shown is **figure 5**. The attenuation length $L$ of the x-rays in the material is determined as $L = 1/\mu$ where $\mu$ is the linear attenuation coefficient of the x-rays in the detector material. The depth of penetration of the x-rays is determined as $z = L \sin\theta$ where $\theta$ is the tilt angle. The geometrically expanded size of the beam due to the tilted angle illumination is determined as $P = a/\sin\theta$, where $a$ is the original size of the beam.

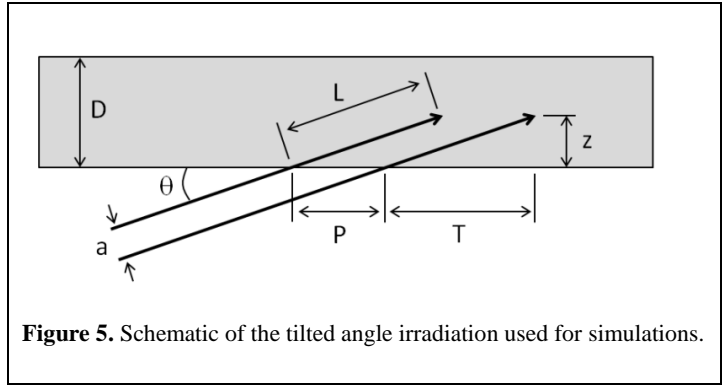

**Figure 5.** Schematic of the tilted angle irradiation used for simulations.

The beam attenuation profile in the direction parallel to the detector surface is determined as

$$\frac{dN}{dx} = \frac{N_0 \mu}{\cos\theta} e^{-\frac{\mu}{\cos\theta} x} \qquad (2)$$

This attenuation profile determines the lateral blurring of the signal due to the tilted angle geometry. The simulations were performed using tilt angle of 6° which provides beam expansion by a factor of 10, but in practice the tilt angle can be adjusted depending on the imaging task.

The simulation steps included (1) expanding the beam and projecting it to the detector surface, (2) 1D convolution of the projected beam by attenuation profile of the beam, (3) 2D convolution with detector response with FWHM including all blurring components, and (4) 1D scaling back (demagnification) of the data by the same beam expansion factor to get non-deformed shape of the imaged object.

No statistical noise was included in current simulations because the purpose was to evaluate effect of the tilted angle on spatial resolution not on signal to noise ratio. However, tilted angle irradiation may have an effect on quantum efficiency and signal to noise ratio, and this will be discussed separately.

*2.2.2. Direct-conversion Si detector based on back illuminated CCD and CMOS arrays*

For some x-ray microscopy applications CCD and CMOS image cameras can be used as direct conversion x-ray detectors. In these cases the scintillator and magnifying optical lens are not used and other mechanisms of magnifications such as x-ray optics based on Fresnel zone plates are used. For example, back illuminated (back thinned) CCD cameras have been used as direct conversion detectors for synchrotron beam-based soft x-ray microscopy at 500eV (Schneider et al. 2002; Carzaniga et al. 2014; Harkiolaki et al. 2018) and 707eV (Chao et al. 2009) photon energies.

Substantial magnification effect and improved spatial resolution can be achieved using tilted angle irradiation of direct conversion CCD/CMOS imagers without using any other magnification mechanisms. The mechanism of the resolution improvement and its evaluation is similar to that for the scintillation detector described in previous sections.

Additional advantage of the tilted angle method with back illuminated CCD and CMOS is that it decreases the depth of penetration of the x-rays into sensor material which is important for preventing radiation damage of the underlying sensor electronics. The negative effect of the radiation on the sensor is cumulative and depends on total x-ray exposure and energy of the x-ray photons. For example, it is recommended that for long term stability against radiation damage the back-illuminated CCD should not be used at photon energies higher than 800eV.

The key energies to be considered are 500eV that is currently used for soft x-ray microscopy, 800eV that is the highest energy can be used with back illuminated CCD, and 2.5keV that is the desirable energy for soft x-ray microscopy. The x-ray attenuation lengths at these energies are 0.43μm, 1.44μm, and 2.72μm, respectively (**figure 4**). Although normal irradiation with 2.5keV x-rays would damage the CCD, we can safely use 2.5keV x-rays in tilted angle configuration. For example, if the 2.5keV x-rays irradiate the CCD detector at 26º tilt angle, the depth of penetration would decrease from 2.72μm in normal irradiation to 1.19μm at 26º tilt angle. On the other hand, the 1.19μm penetration depth corresponds to the x-ray energy of 750eV in normal irradiation mode, which is safe for using CCD.

The back illuminated CCD with a pixel size of 10x10μm$^2$ and sensor thickness of 10μm was used in simulation studies. These CCDs are manufactured with 1024x1024 and 2048x2048 pixel arrays (Princeton Instruments). The simulation procedure was similar to that described above for scintillator based imaging system. The tilt angles were 6º and 26º, and the detector response function was a Gaussian function with 10μm FWHM. Although the detector pixels are square-shaped, in practice the edges of the response function are blurred due to various effects including charge sharing, imperfect pixel shapes, etc. The magnitude of this blurring may vary for different systems, and for this reason we used Gaussian shaped response function with the same FWHM as the square-shaped pixel size.

*2.2.3. Phantoms*

We used digital phantoms for qualitative and quantitative evaluation of the effect of tilted angle irradiation on spatial resolution (**Figure 6**). The concept of the phantom was similar to that used in previous experimental work where the x-ray microscopy system was evaluated by imaging copper interconnect structures of electronic circuits (Neuhausler et al. 2003). The phantom included a set of parallel strips with 0.1-0.5 μm widths and the gaps in between were the same as strip widths. The phantom included also square pads with sizes of 0.1-0.5μm and same gaps in between. The above phantom was modified for testing direct conversion CCD detector. It had the same strip and pad configurations but the sizes of these components were larger by a factor of 40, being in the range of 4-20 μm. The 40 times difference as compared to the phantom for the scintillation detector reflects the optical magnification of the scintillation detector that the direct conversion detector does not have.

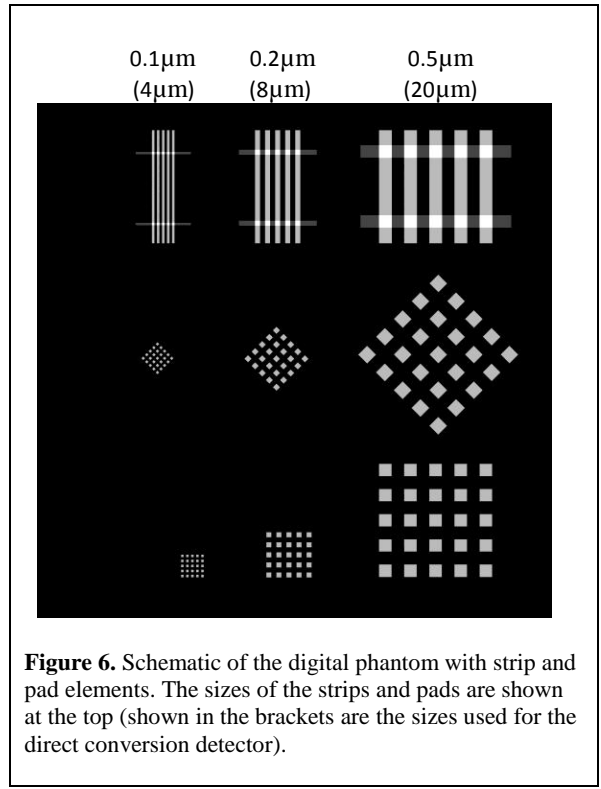

**Figure 6.** Schematic of the digital phantom with strip and pad elements. The sizes of the strips and pads are shown at the top (shown in the brackets are the sizes used for the direct conversion detector).

## 3. Results

**Figure 7** shows the detector response function in normal and tilted angle irradiation modes for indirect detector. In normal irradiation mode the response function has a Gaussian shape with 0.7μm FWHM. The response function in normal irradiation mode is independent of x-ray energy because at 2-5keV energies the x-ray interaction component of the blurring is much smaller than the other components. In tilted angle case, as opposed to normal irradiation, the response functions are asymmetric and dependent on the energy. This is due to the lateral penetration of the x-rays when tilted angle is used, and response is broadened as the x-ray energy is increased. Although in tilted angle case the lateral blurring due to x-ray penetration was combined with 0.7μm intrinsic detector blurring, the intrinsic blurring did not have a strong effect because it was applied to the expanded image which then was scaled back (demagnification) decreasing effect of the 0.7μm blurring by 10 times.

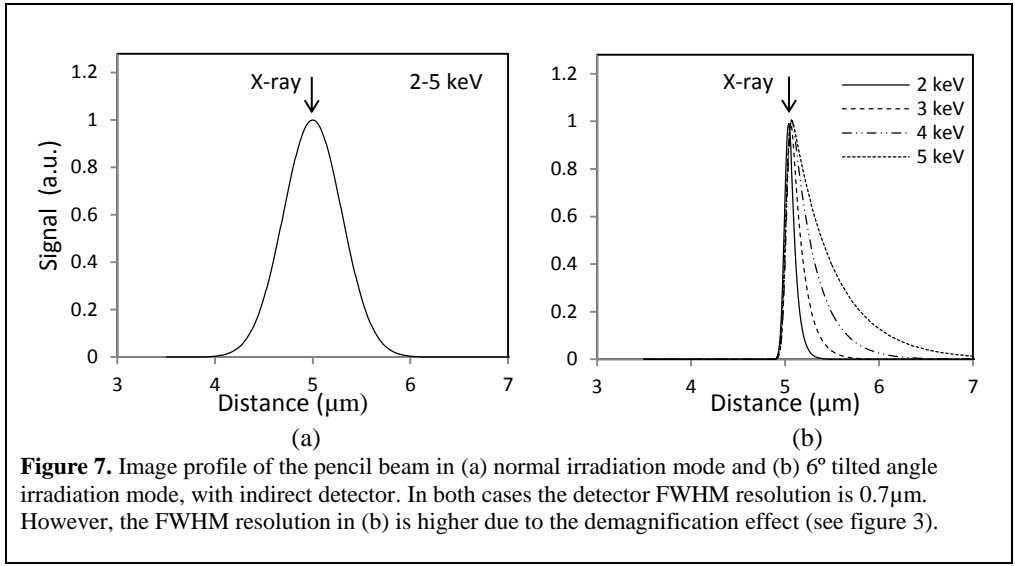

**Figure 7.** Image profile of the pencil beam in (a) normal irradiation mode and (b) 6º tilted angle irradiation mode, with indirect detector. In both cases the detector FWHM resolution is 0.7µm. However, the FWHM resolution in (b) is higher due to the demagnification effect (see figure 3).

**Figure 8** shows the attenuation profiles of the two pencil beams of the x-rays in 90º (normal) irradiation mode, and corresponding image profiles showing blurring due to detector response. The gaps between the beams were 0.5µm and 1µm, respectively. As can be seen, the 0.7µm FWHM of the detector response function results in complete merging of the signals for 0.5µm gap case, and in the 1µm gap case the two beams are barely separated.

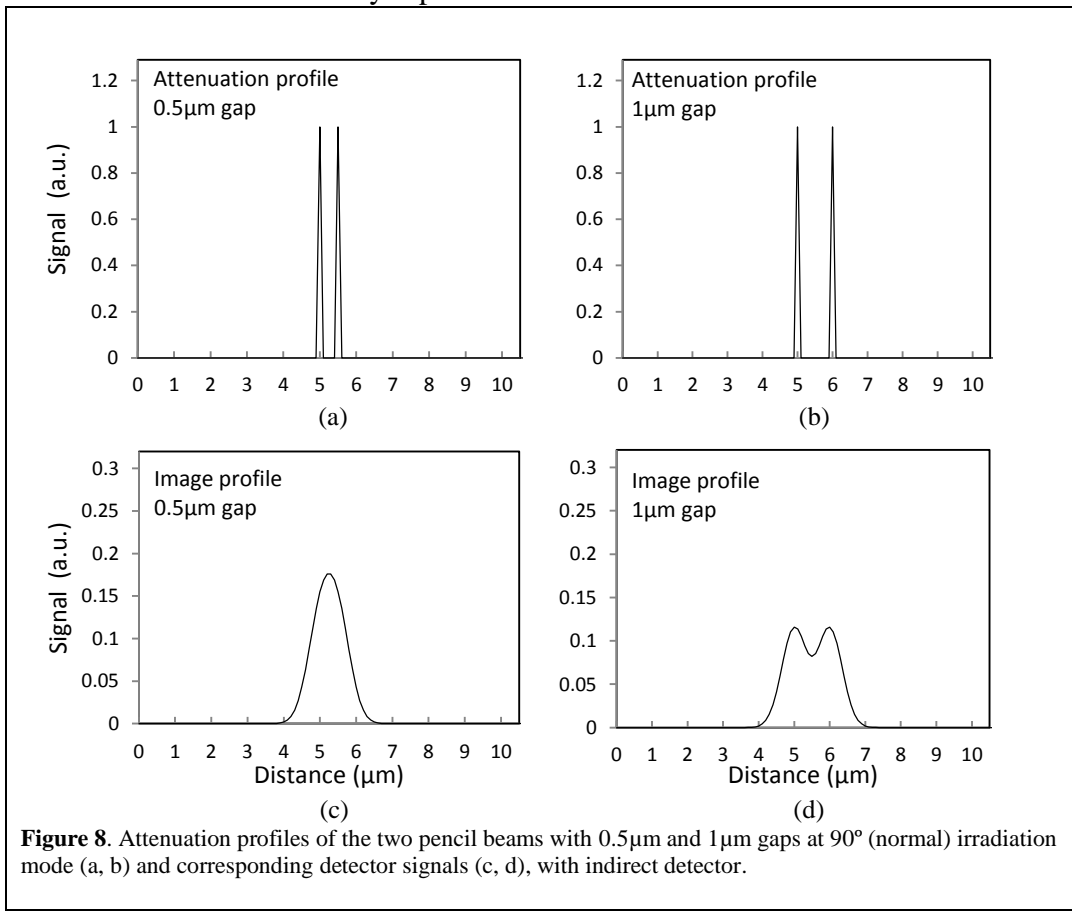

**Figure 8**. Attenuation profiles of the two pencil beams with 0.5µm and 1µm gaps at 90º (normal) irradiation mode (a, b) and corresponding detector signals (c, d), with indirect detector.

The **figure 9** shows the image profiles of the two pencil beams with 0.5μm gaps and at 6º tilt angle. The profiles were simulated for 2keV (a, b, c) and 5keV (d-m) x-ray energies. For the images (g, k, m) the thickness of the scintillator was decreased from 5μm to 0.25μm so that the overlaps of the attenuation profiles seen in (d, e, f) are eliminated. The figures (a, d, c) represent attenuation profiles, (b, e, k) attenuation profiles after 0.7μm detector blurring, and the (c, f, m) are the scaled back (demagnified) versions of (c, f, m) to provide non-deformed shape of the object.

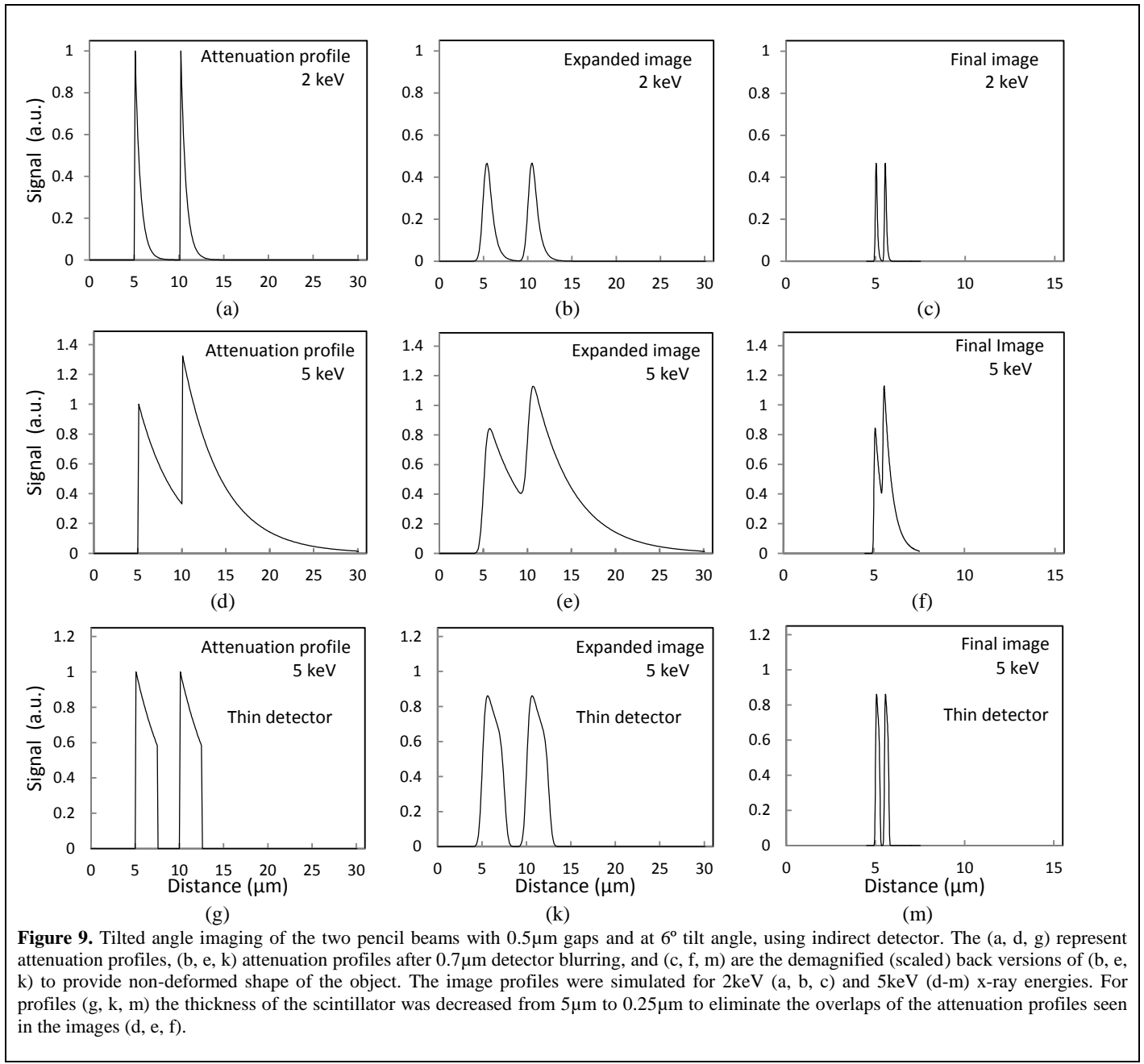

**Figure 9.** Tilted angle imaging of the two pencil beams with 0.5μm gaps and at 6º tilt angle, using indirect detector. The (a, d, g) represent attenuation profiles, (b, e, k) attenuation profiles after 0.7μm detector blurring, and (c, f, m) are the demagnified (scaled) back versions of (b, e, k) to provide non-deformed shape of the object. The image profiles were simulated for 2keV (a, b, c) and 5keV (d-m) x-ray energies. For profiles (g, k, m) the thickness of the scintillator was decreased from 5μm to 0.25μm to eliminate the overlaps of the attenuation profiles seen in the images (d, e, f).

**Figure 10** shows how the indirect detector resolution evolves in tilted angle geometry when the x-ray energy is increased. The image profiles of the two pencil beams with 0.5μm gaps and 2keV, 3keV, 4keV, and 5keV energies, at 6º tilt angle show the increased signal overlap and deterioration of the resolution as x-ray energy increases.

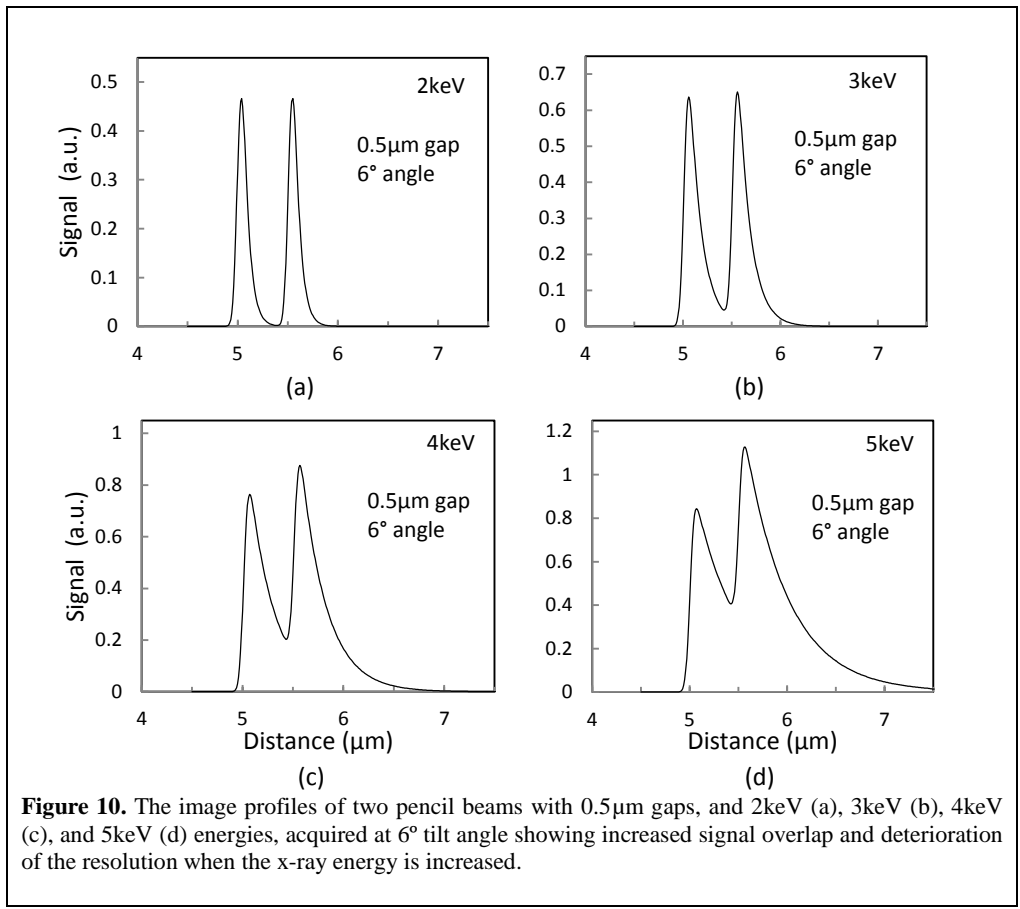

**Figure 10.** The image profiles of two pencil beams with 0.5μm gaps, and 2keV (a), 3keV (b), 4keV (c), and 5keV (d) energies, acquired at 6º tilt angle showing increased signal overlap and deterioration of the resolution when the x-ray energy is increased.

**Figure 11** shows simulated images of the digital phantom with strip and pad elements with indirect detector. The images were simulated with 90º (normal) irradiation, and 6º tilted angle irradiation, at 2keV, 3keV, 4keV, and 5keV x-ray energies. The intrinsic resolution of the detector was 0.7μm FWHM in normal and tilted angle irradiations. As can be seen, normal irradiation exhibits 2D blurring of the image while tilted angle irradiation provides high resolution in one dimensions and similar blurring in the second direction. In tilted angle geometry, turning the pad arrays by 45º provides "sharing" the resolution in two dimensions.

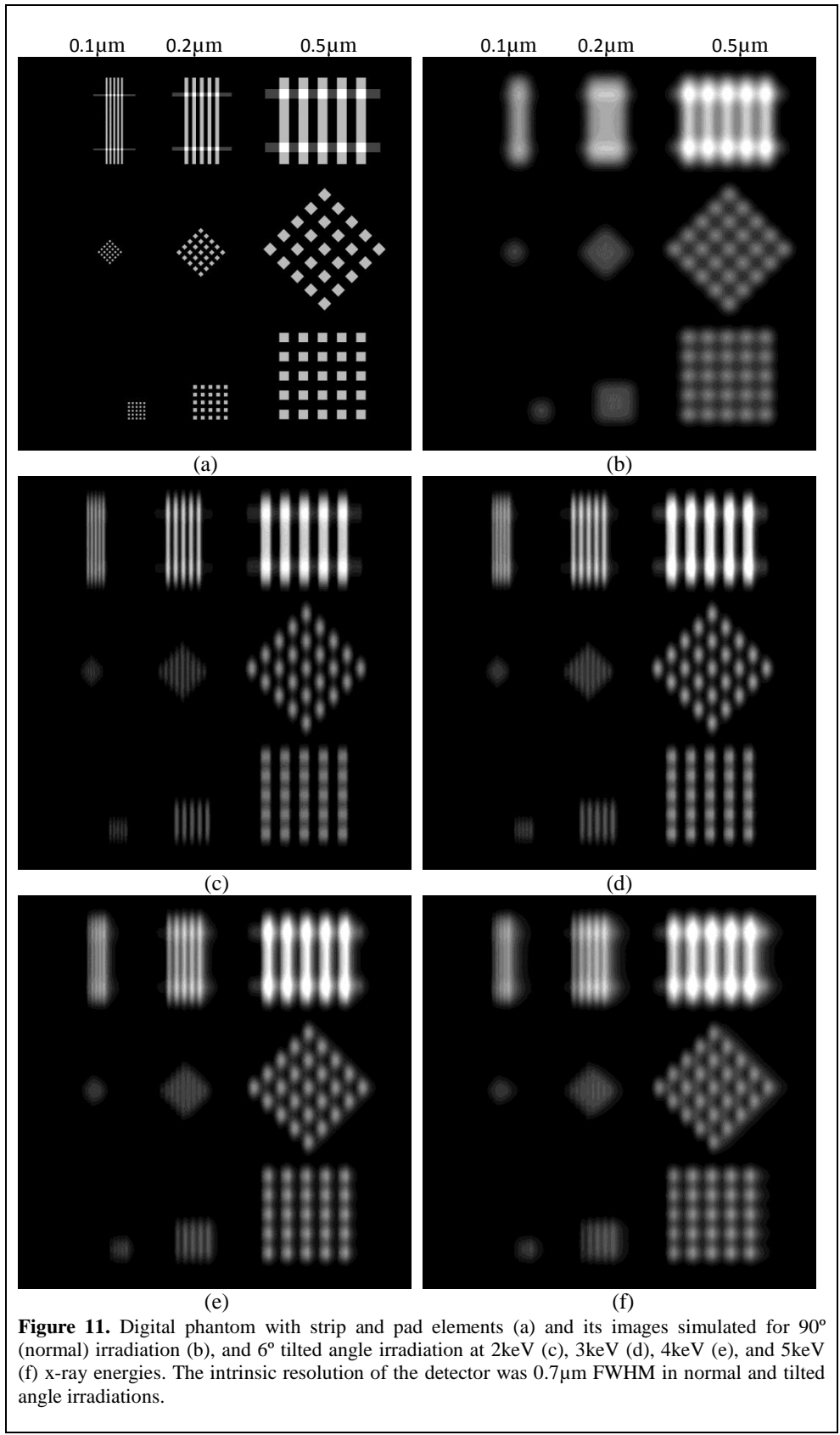

**Figure 11.** Digital phantom with strip and pad elements (a) and its images simulated for 90º (normal) irradiation (b), and 6º tilted angle irradiation at 2keV (c), 3keV (d), 4keV (e), and 5keV (f) x-ray energies. The intrinsic resolution of the detector was 0.7µm FWHM in normal and tilted angle irradiations.

**Figure 12** presents the image profiles of the strip elements of the phantom showing how the tilted angle irradiation improves the resolution. The resolution can be quantified using magnitude of the contrast transfer (CT) at different spatial frequencies. While the normal irradiation completely blurs out the images of 0.1μm and 0.2μm strips the tilted irradiation allows resolving them. However, the tail pile-up effect is evident with tilted angle geometry and it is strongly pronounced at higher energies and/or for finer strips.

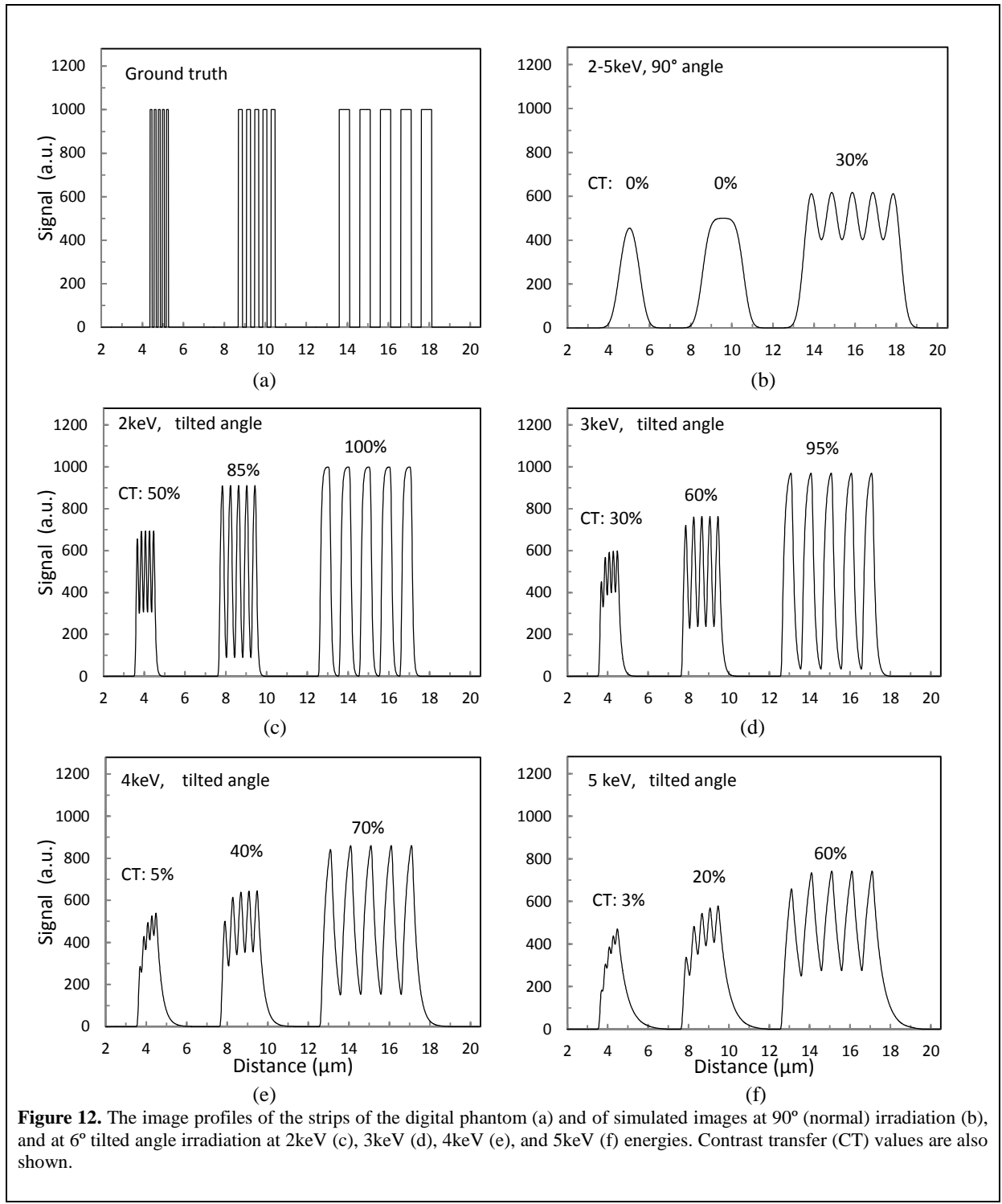

**Figure 12.** The image profiles of the strips of the digital phantom (a) and of simulated images at 90° (normal) irradiation (b), and at 6° tilted angle irradiation at 2keV (c), 3keV (d), 4keV (e), and 5keV (f) energies. Contrast transfer (CT) values are also shown.

**Figure 13** demonstrates the effect of decreasing indirect detector thickness to eliminate tail pileup in tilted angle irradiation geometry. As can be seen, decreasing the scintillator thickness from 5µm to 0.25µm allows for lowering signal tail pileups and complete separation of the signals from two pencil beams with 0.5µm gap in between. Although the scintillator thickness was decreased to 0.25µm, the x-ray attenuation length was 2.5µm due to the tilted angle geometry.

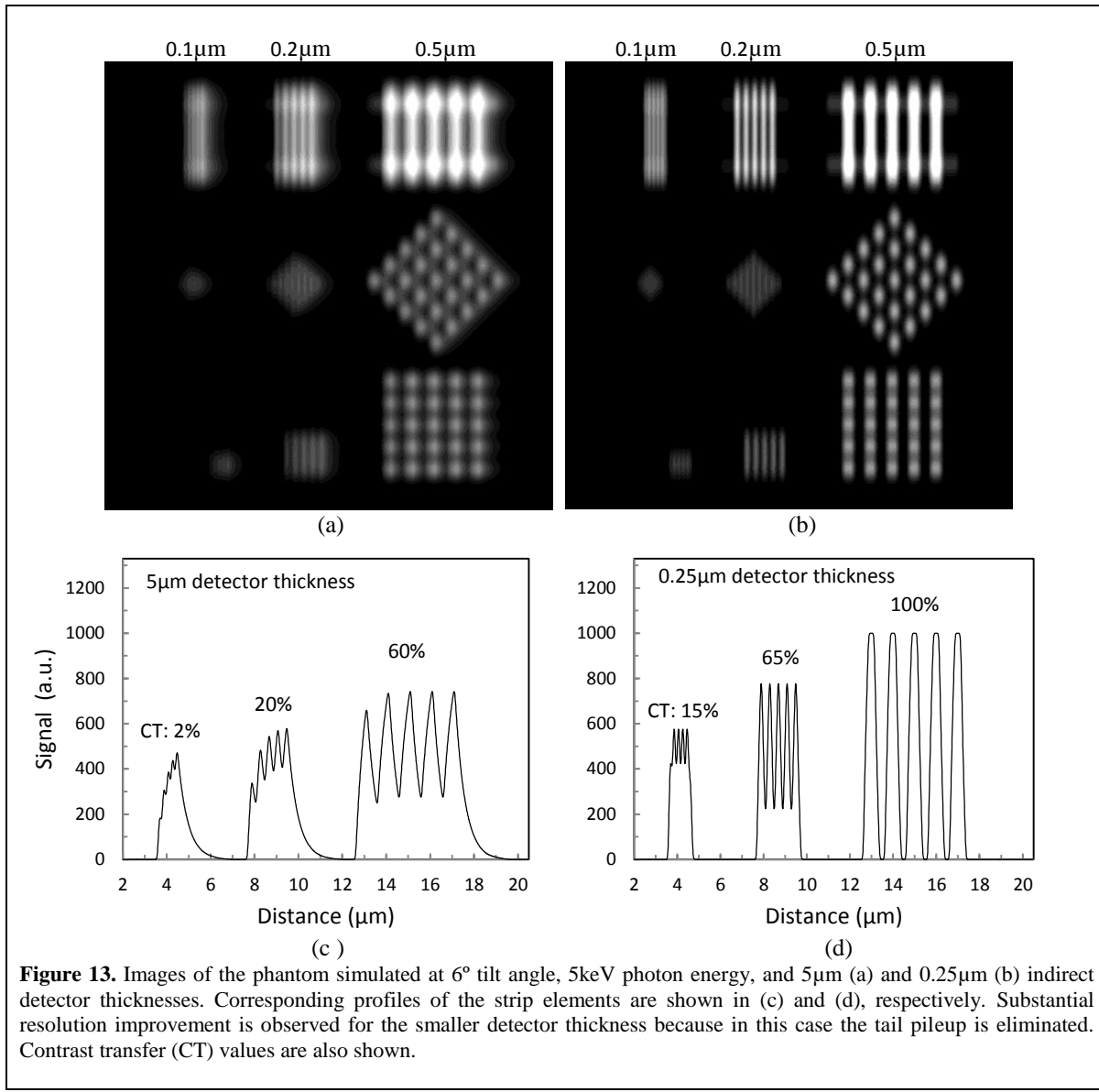

**Figure 13.** Images of the phantom simulated at 6º tilt angle, 5keV photon energy, and 5µm (a) and 0.25µm (b) indirect detector thicknesses. Corresponding profiles of the strip elements are shown in (c) and (d), respectively. Substantial resolution improvement is observed for the smaller detector thickness because in this case the tail pileup is eliminated. Contrast transfer (CT) values are also shown.

**Figure 14** shows the images of the second phantom designed for direct-conversion back illuminated CCD image camera with 10x10 µm pixel size and Si sensor. The sizes of the resolution elements are 40 times larger than that of previous phantom. The images acquired at 2.5keV x-ray energy with normal irradiation, and with tilted angle irradiations at 26º and 6º tilt angles show that the tilted angle irradiation provides substantial increase of the spatial resolution.

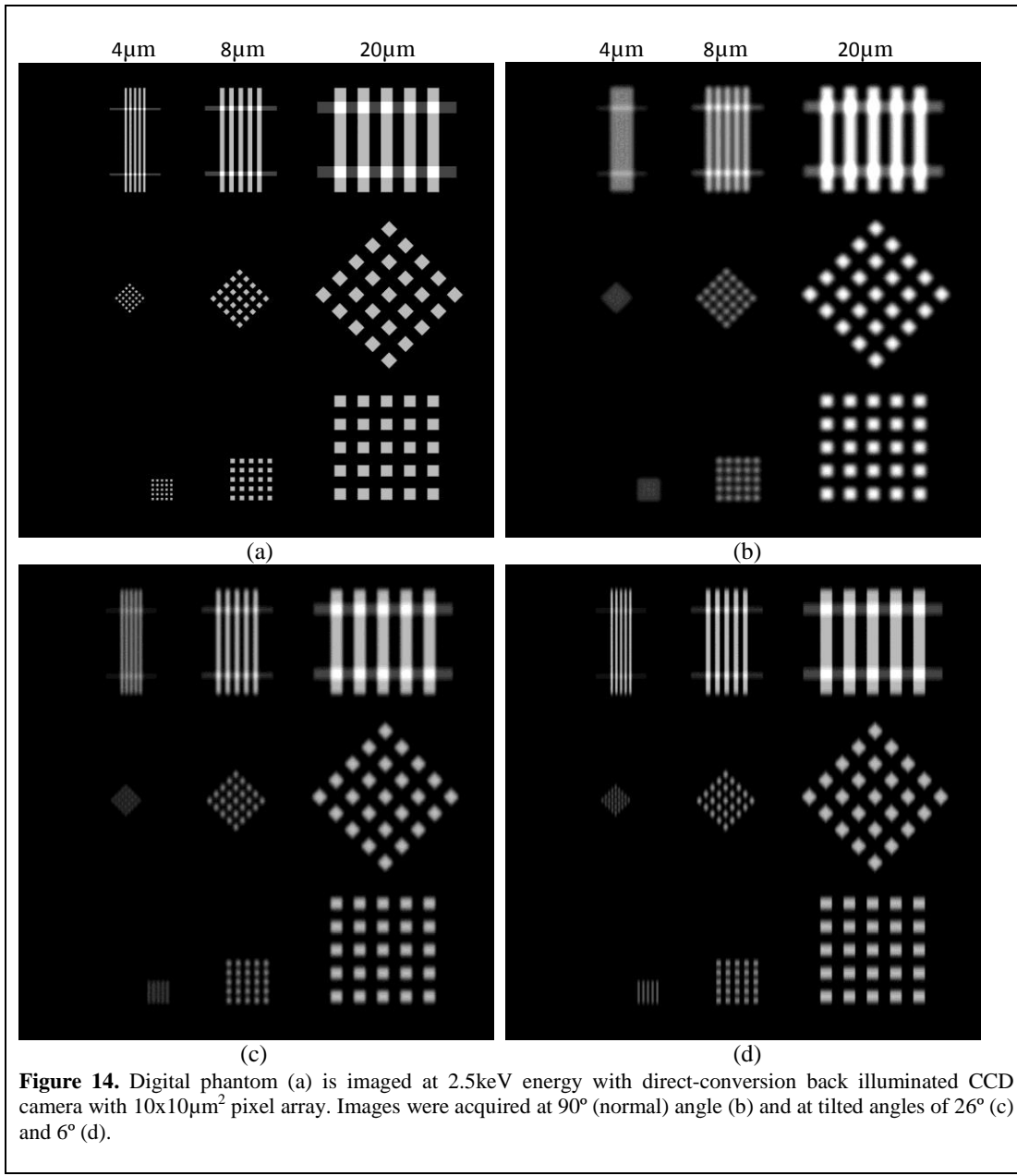

**Figure 14.** Digital phantom (a) is imaged at 2.5keV energy with direct-conversion back illuminated CCD camera with 10x10μm$^2$ pixel array. Images were acquired at 90º (normal) angle (b) and at tilted angles of 26º (c) and 6º (d).

**Figure 15** shows clear advantage of the tilted angle irradiation when it is applied to direct-conversion back illuminated CCD detectors. While the normal irradiation provides approximately 0% and 50% modulation transfer for 4μm and 8μm width strips, respectively, the 26º tilt angle provides 50% and 90% modulation transfer, respectively, for the same strip elements. Decreasing tilt angle to 6º provides 100% modulation transfer for both of the above strip elements. In addition to the improved resolution, with tilted angle irradiation the x-rays are absorbed in shallow depths of the Si sensor which prevents the radiation induced damages of the underlying electronic circuits.

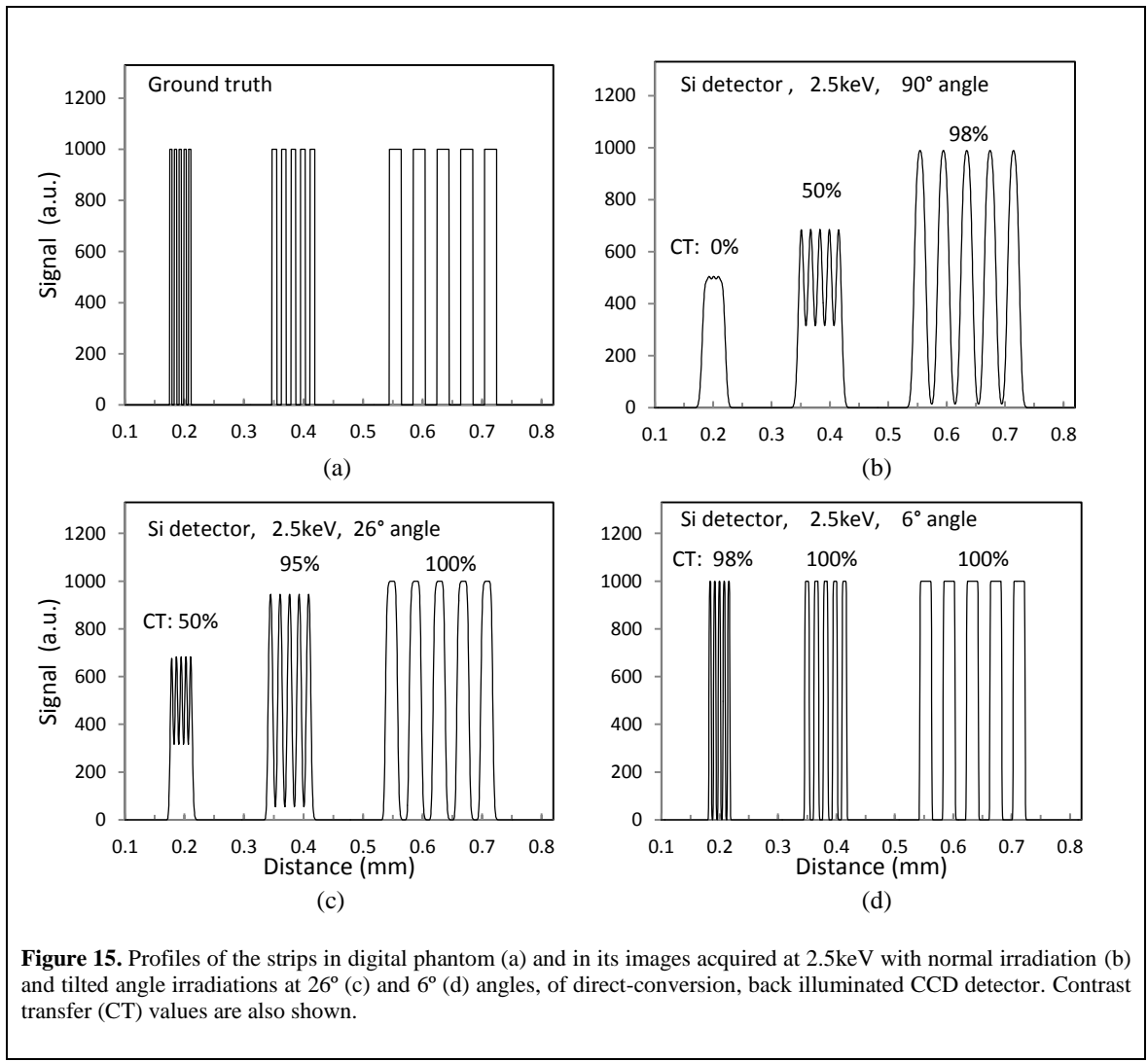

**Figure 15.** Profiles of the strips in digital phantom (a) and in its images acquired at 2.5keV with normal irradiation (b) and tilted angle irradiations at 26º (c) and 6º (d) angles, of direct-conversion, back illuminated CCD detector. Contrast transfer (CT) values are also shown.

## 4. Discussion and conclusion

Potential advantages of the tilted angle irradiation in x-ray microscopy have been evaluated by simulation studies and reported in this work. Along with the advantages, potential limitations of the tilted angle irradiation as well as the methods for addressing them are also reported.

Substantial improvement of the spatial resolution with tilted angle method is achieved for low energy x-rays with energies up to 5keV. Soft x-ray microscopy using this particular energy range has important applications where improving spatial resolution is of interest (Schneider et al. 2002; Carzaniga et al. 2014; Harkiolaki et al. 2018). At higher x-ray energies the signal overlap due to the tail pileup effect decreases spatial resolution (see **figures 1 and 9**). This problem can be addressed by decreasing the detector thickness, although this will decrease also quantum efficiency. For example, the tail pileup effect at 5keV energy and 6º tilt angle was eliminated by decreasing scintillator thickness from 5μm to 0.25μm (**figure 9**). The 0.25μm detector thickness corresponds to 2.5μm x-ray path length in the detector at 6º tilt angle. For LuAG scintillator used in our study the 2.5μm thickness of the scintillator absorbs 43% of 5keV x-ray photons. For the x-ray energies higher than 5keV the photon absorption with 0.25µm thick scintillators decreases and average detection efficiency for 5-12keV photons at 6º tilt angle is 20%. Notice also that for the detectors currently used in x-ray microscopy the collection efficiency of the scintillation light by optical lens is less than 1% (Koch et al. 1998; Martin and Koch 2006). Tilted angle configuration could improve light collection efficiency by direct collection of the light without having passed it through scintillator substrate. Therefore, efficiency loss due to the decreased scintillator thickness can be partially or fully compensated by increased light collection efficiency in tilted angle mode.

It appears from the **figure 1** that the tilted angle irradiation method could work best when parallel x-ray beams are used. However, some soft x-ray microscopy systems use nonparallel x-ray beams that are focused at the detector surface using objective lenses such as Fresnel zone plates. If tilted angle detector is used in these systems then the distance between the lens and detector surface will vary and focal point will blur from center toward the edges of the detector. On the other hand, the diameters of the x-ray objective lenses are very small (few tens of microns) and lens-to-detector distances are large (few meters) (Schneider et al. 2002; Neuhausler et al. 2003; Carzaniga et al. 2014; Harkiolaki et al. 2018) so that the variation of the lens-to-detector distance due to the tilted angle alignment of the detector results in negligible focal spot blurring. For example, the x-ray lens based on Fresnel zone plate used in the work (Neuhausler et al. 2003) had approximately 52µm diameter and the lens-to-detector distance was 2660mm. If we use, for example, a detector with 13x13mm$^2$ area and tilt angle of 6º then the largest cross-sectional blurring of the focal spot at the near and far edges of the detector will be 130nm. This 130nm size will be further expanded in 1D by a factor 10 due to the tilted geometry, reaching 1.3µm, which is still much smaller than the typical detector pixel size of 10µm.

Although tilted angle method provides resolution improvement in one dimension, the method could still be very useful. As can be seen from **figure 11**, the images of the square pad arrays are much better visualized when they are turned by 45º because in this case the improved 1D resolution is "shared" and extended to the second dimension helping better visualize 2D structures. Also, for some applications such as fan beam CT, improved 1D resolution is required to achieve highest resolution in XY (slice) plane while the resolution requirements in Z (slice thickness) direction can be relaxed. It is also possible to turn the tilted angle detector (or imaged object) around axial direction and acquire data at different rotation angles providing more high resolution data at different rotation angles. With sufficient angular sampling it can be possible to reconstruct pure 2D data with the highest resolution provided by tilted angle detector.

It should be noticed also that because tilted angle configuration expands the beam in one dimension the detector field of view may need to be extended in one dimension. On the other hand, higher resolution is often required for imaging smaller objects and the larger detector may not be necessary in many cases. Also, there are detectors that can be tiled up to provide larger sizes in one dimension and these detectors are well suited for tilted angle irradiation.

Additional advantage of the tilted angle method is that it can prevent back illuminated CCD and CMOS detectors from getting radiation damage. The tilted angle method decreases the depth of penetration of the x-rays into sensor material which is important for preventing radiation damage of the underlying sensor electronics (see **section 2.2.2**). For example, it is recommended that for long term stability against radiation damage the back-illuminated CCD should not be used at photon energies higher than 800eV. However, these CCDs can be used at higher than 800eV energies if tilted angle geometry is used. As described in the above section 2.2.2, the depth of penetration for 2.5keV x-rays at 26º tilt angle is the same as for 750eV x-rays at normal irradiation, and therefore, the back illuminated CCD can be used to measure 2.5keV x-rays at 26º tilt angle.

Overall, the results and discussions show that the tilted angle irradiation approach can be a useful tool for increasing spatial resolution in the cases when further increasing resolution is limited by the fundamental physical factors.